\documentclass[aps,pre,twocolumn,groupedaddress,superscriptaddress,showpacs]{revtex4-1}
\usepackage{graphicx}
\usepackage[caption=false]{subfig}
\usepackage{latexsym}
\usepackage{amsmath}
\usepackage[makeroom]{cancel}
\usepackage{relsize}
\usepackage{letltxmacro}
\usepackage{amssymb}
\usepackage{epstopdf}
\usepackage{ulem}
\usepackage{textcomp} % for the use of \textdegree

\usepackage[usenames]{color}

\newcommand{\sn}[1]{{{#1}}}
\LetLtxMacro{\oldsqrt}{\sqrt} % makes all sqrts closed
\renewcommand{\sqrt}[1][\ ]{%
  \def\DHLindex{#1}\mathpalette\DHLhksqrt}
\def\DHLhksqrt#1#2{%
  \setbox0=\hbox{$#1\oldsqrt[\DHLindex]{#2\,}$}\dimen0=\ht0
  \advance\dimen0-0.2\ht0
  \setbox2=\hbox{\vrule height\ht0 depth -\dimen0}%
  {\box0\lower0.71pt\box2}}

\def\be{\begin{equation}}
\def\ee{\end{equation}}
\def\bea{\begin{eqnarray}}
\def\eea{\end{eqnarray}}
\def\ba{\begin{array}}
\def\ea{\end{array}}

\begin{document}

\title{Normal stresses in semiflexible polymer hydrogels}
\author{M.\ Vahabi}\affiliation{Department of Physics and Astronomy, Vrije Universiteit, Amsterdam, The Netherlands}
\author{Bart E. Vos}\affiliation{AMOLF, Department of Living Matter, 1098 XG Amsterdam}%
\author{Henri C. G. de Cagny}\affiliation{Institute of Physics, University of Amsterdam, Amsterdam, The Netherlands}
\author{Daniel Bonn}\affiliation{Institute of Physics, University of Amsterdam, Amsterdam, The Netherlands}
\author{Gijsje H. Koenderink}\affiliation{AMOLF, Department of Living Matter, 1098 XG Amsterdam}%
\author{F.\ C.\ MacKintosh}\affiliation{Department of Physics and Astronomy, Vrije Universiteit, Amsterdam, The Netherlands}\affiliation{Department of Chemical \& Biomolecular Engineering, Rice University, TX 77005 Houston, USA}\affiliation{Center for Theoretical Biological Physics, Rice University, TX 77030 Houston, USA}\affiliation{Departments of Chemistry and Physics \& Astronomy, Rice University, TX 77005 Houston, USA}
\date{\today}

\begin{abstract}
Biopolymer gels such as fibrin and collagen networks are known to develop tensile axial stress when subject to torsion. This negative normal stress is opposite to the classical Poynting effect observed for most elastic solids including synthetic polymer gels, where torsion provokes a positive normal stress. \sn{As shown recently,} this anomalous behavior in fibrin gels depends on the open, porous network structure of biopolymer gels, which facilitates interstitial fluid flow during shear and can be described by a  phenomenological two-fluid model with viscous coupling between network and solvent. Here we extend this model and develop a microscopic model for the individual diagonal components of the stress tensor that determine the axial response of semi-flexible polymer hydrogels. This microscopic model predicts that the magnitude of these stress components depends inversely on the characteristic strain for the onset of nonlinear shear stress, which we confirm experimentally by shear rheometry on fibrin gels. Moreover, our model predicts a transient behavior of the normal stress, which is in excellent agreement with the full time-dependent normal stress we measure. 
\end{abstract}
\pacs{}
\keywords{}
\maketitle

\section{Introduction: Normal stresses in semiflexible polymer gels}
\begin{figure}[htbp]
\includegraphics[width=0.7\columnwidth]{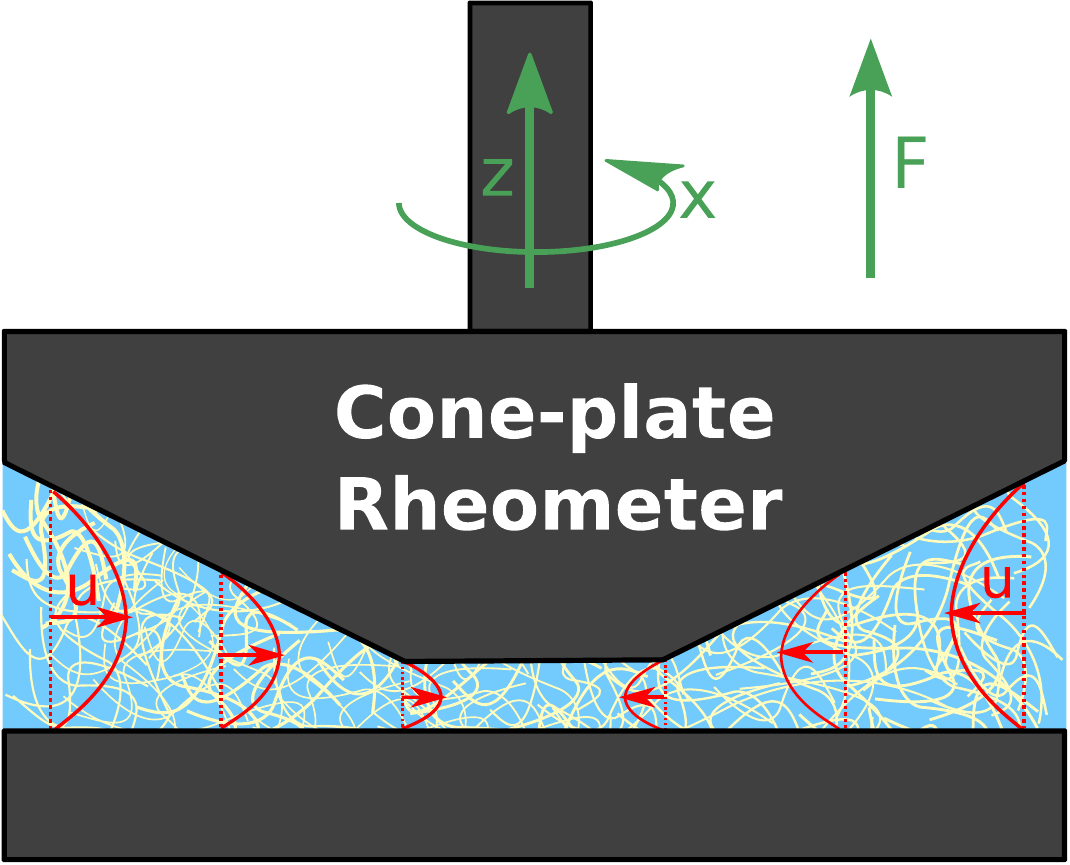}
\caption{
(color online) Schematic representation of a semiflexible polymer hydrogel in a cone-plate rheometer and of the two-fluids model, which allows for an inward, radial displacement of the network ($u$: red inward arrows) relative to the radially stationary solvent upon the application of a shear stress by rotation of the cone. This geometry also defines our coordinates, with $x$ corresponding to the azimuthal (shear) direction and $z$ the axial (gradient) direction.}
\label{fig:1}
\end{figure}

A little over a hundred years ago, Poynting demonstrated in a series of experiments that most elastic materials elongate axially when subject to torsion, as in the case of a twisted wire or elastic rod \cite{Poynting1909,PoyntingRubber}. Fundamentally, this {\it Poynting effect} is a manifestation of nonlinear elasticity, since symmetry requires that elongation also occurs for torsion of the opposite sign, unless the material is chiral. Being a nonlinear effect, the degree of elongation can be expected to vary initially quadratically in the torsional strain, meaning that the effect tends to be weak unless the strain is large. The Poynting effect is also commonly observed in torsional rheometry of soft materials. This is illustrated schematically in Fig.\ \ref{fig:1} for a polymer gel, where a positive axial force $F$ generally develops if the sample height is fixed. Again, this normal stress is generally quadratic in strain and weak except at large strain \cite{Barnes1989}. 

Biopolymer gels, such as those that occur naturally in living cells and tissues, represent a class of materials that have been studied extensively, both theoretically and experimentally, particularly for their highly nonlinear elastic properties \cite{Janmey1994,Gardel2004,Storm2005,Kasza2007,Broedersz2014}. Such systems have been shown to exhibit, for instance, 10-fold or more stress-stiffening when subject to even small strains as low as a few percent, which suggests similarly strong normal stress effects. 
It was recently shown that normal stresses are, indeed, anomalously large for a wide range of biopolymer gels \cite{Janmey2007,Kang2009}. But, more surprisingly, it was also shown that the sign of the normal stress was {\it opposite} to that of synthetic polymer gels: collagen, fibrin and other biopolymer gels tend to {\it contract} axially when subject to torsion. 
It was argued theoretically in Refs.\ \cite{Janmey2007,Kang2009,Conti2009} that the negative sign of the observed normal stress was the result of compressibility of the network.

Perhaps counterintuitively, the measured axial force in an experiment such as the one sketched in Fig.\ \ref{fig:1} is not a direct measure of the diagonal axial stress component $\sigma_{zz}$ in the stress tensor. There is an additional contribution due to the azimuthal term $\sigma_{xx}$ that arises from hoop stress in the torsional geometry \cite{Venerus2007}. It is not generally possible to directly measure individual diagonal stress components such as $\sigma_{zz}$ by conventional rheometry. 
This is due to the fact that the diagonal terms in the stress tensor also involve pressure, which can vary within the sample. Strictly speaking, for incompressible materials, this means that only normal stress differences, such as $N_1=\sigma_{xx}-\sigma_{zz}$, can be measured by an experiment such as the one illustrated in Fig.\ \ref{fig:1}. 
The first of these terms arises from the curved azimuthal streamlines that give rise to hoop stresses proportional to $\sigma_{xx}$ in the (inward) radial direction. 
In an incompressible medium, no radial displacement of the gel is possible, and a radial pressure gradient develops to satisfy force balance. 
The resulting excess pressure (over ambient pressure at the radial boundary) gives rise to the positive contribution to the thrust $F$ measured on the cone. 
In contrast, biopolymer gels such as those of fibrin, with pore sizes in the micrometer range \cite{Okada1985,Pieters2012,Lang2013}, can expel interstitial fluid to relax pressure gradients on experimentally relevant time scales, allowing the network to contract upon shearing \cite{heussinger2007nonaffine,Tighe2013}. 
For torsional rheology, this effectively leaves the pure axial $\sigma_{zz}$ to dominate the measured thrust \cite{Janmey2007,Kang2009,Conti2009}. 
The thrust is thus expected to change sign from positive to negative over a timescale governed by the porosity of the network that allows it to move relative to the incompressible solvent.

\begin{figure}[b]
\centering
%\vspace{.1cm}
\includegraphics[width=8cm]{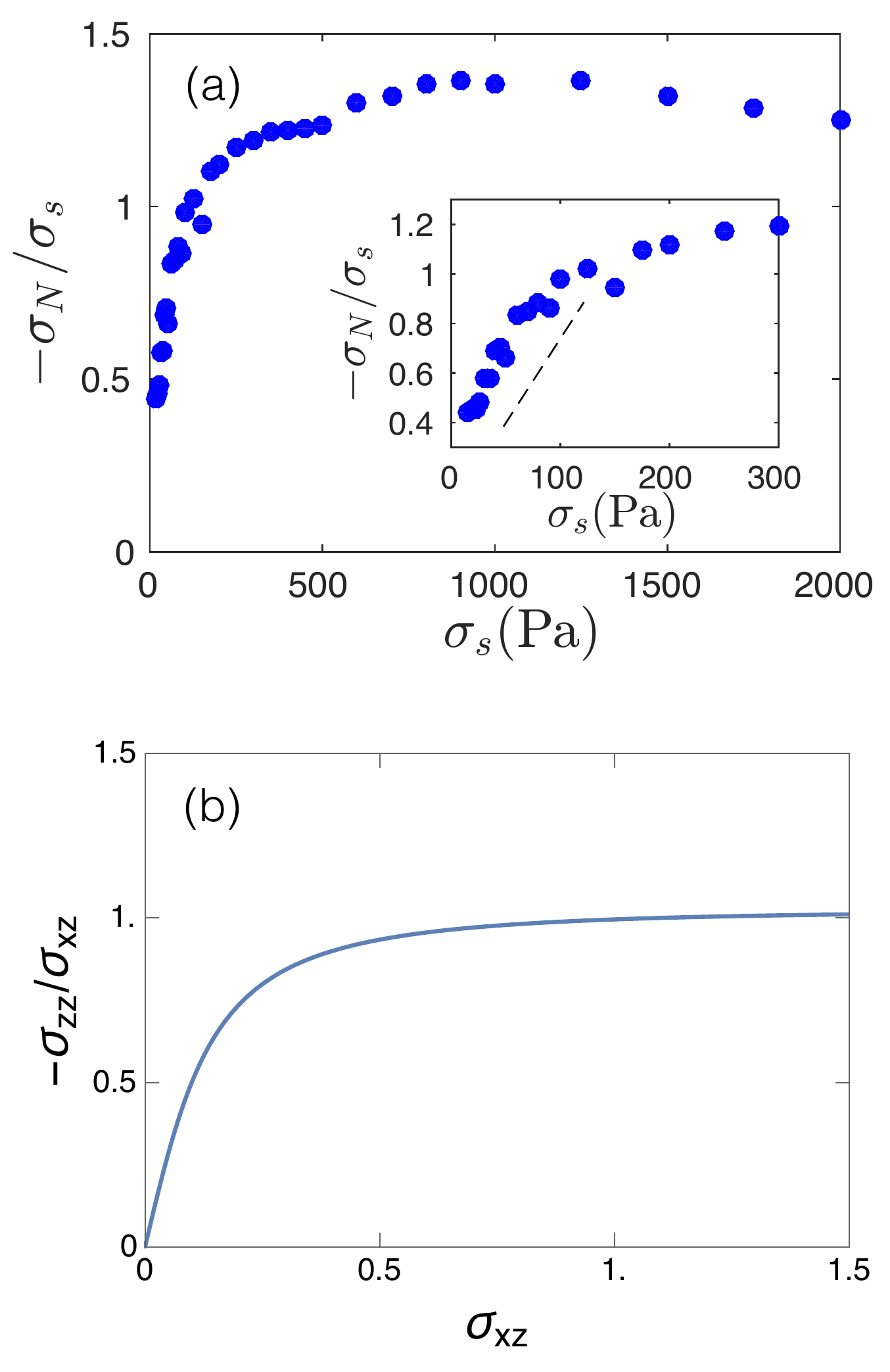}
\caption{(a) Magnitude of the normal stress $\sigma_N=\frac{2F}{\pi R^2}$, as measured by the rheometer from the thrust $F$ (see Fig.\ \ref{fig:1}), normalized by the shear stress $\sigma_s$ is plotted versus the shear stress for a fibrin gel polymerized at 22\textdegree C ($G^{'} = 963$ Pa, $\nu=1$ Hz) . Following an initial approximately linear regime (inset), a saturation of the ratio $|\sigma_N/\sigma_s|$  to a value of order unity is found. (b) The dimensionless ratio of the axial stress $\sigma_{zz}$ and shear stress $\sigma_{xz}$, plotted vs shear stress (arbitrary units), as predicted for a semiflexible gel in the limit of large persistence length compared with the network mesh size (see Sec.\ \ref{C}) \cite{Janmey2007,Kang2009}. }
\label{fig:NormStress}
\end{figure}

We recently reported a direct observation of this predicted change of sign of the normal stress from positive to negative in both fibrin networks and synthetic polyacrylamide (PAAm) hydrogels \cite{DeCagny2016}. 
Moreover, the timescale for this change in sign was shown to depend on the pore size and the elasticity of the network, the solvent viscosity and the gap size of the rheometer. Networks of the blood-clotting protein fibrin were used as model hydrogels, in part because their pore size can be controllably adjusted through the polymerization temperature. The smaller the pore size, the stronger the viscous coupling between the network and the solvent and the longer the characteristic time for reversal of normal stress, which was observed in these experiments. 
The relative strength of the normal to shear stresses was also shown to be larger in magnitude at a given level of strain than for conventional hydrogels such as PAAm. This represents yet another manifestation of the highly nonlinear elastic properties of biopolymer gels. 
In contrast with rubber, where normal and shear stresses become comparable only at strains of order unity, 
both affine-thermal \cite{Janmey2007,Kang2009} and athermal models \cite{Heussinger2007,heussinger2007nonaffine,Conti2009} of semiflexible polymer networks predict that this occurs at small strains $\gamma\simeq10\%$ or less. 
This threshold coincides with the onset strain $\gamma_0$ of nonlinear stiffening in the shear stress $\sigma_{xz}$. Specifically, it is predicted that \cite{Janmey2007,Conti2009}
\begin{equation}
\left|\frac{\sigma_{zz}}{\sigma_{xz}}\right|\sim\frac{\gamma}{\gamma_0} 
\end{equation}
where this ratio saturates to a value of order 1 for $\gamma\gtrsim\gamma_0$, consistent with measurements on fibrin gels shown in Fig.\ \ref{fig:NormStress}.

In Ref.\ \cite{DeCagny2016} we also developed a phenomenological model for the time dependence of the normal stress, \sn{based on the so called two-fluid model of an elastic network that is viscously coupled to a fluid in which the network is embedded \cite{Brochard1977,Milner1993,Gittes1997,Levine2000}. The model in Ref.\ \cite{DeCagny2016} should be generally applicable to flexible or semiflexible polymer gels with a solvent. We showed that this model could account for the observed dependence of the normal stress on porosity and sample geometry for both flexible PAAm and semiflexible fibrin gels.} 
Here, we expand on the model presented in Ref.\ \cite{DeCagny2016}, with full derivations of both transient and time-dependent steady-state evolution of the normal stress. 
Moreover, we calculate \sn{the relevant terms in the stress tensor for semiflexible polymer networks, thereby identifying phenomenological parameters in the prior model.} 
We also present experimental data from fibrin networks polymerized under different conditions for comparison with our model. We find good agreement in both transient and steady-state regimes. 
 
The paper is organized as follows: Section \ref{Model} describes the two-fluid model. Section \ref{C} describes the calculation of different stress components used in the model. Section \ref{Exp} explains the experimental methodology. In Section \ref{Results}, we present and discuss our results.

\section{Two-fluid Model and stress relaxation in gels}\label{Model}
%\subsection{Two-fluid model and relaxation of hoop stress}\label{B}
When a viscoelastic gel is sheared in a cone and plate rheometer, tension tends to build up along the streamlines, giving rise to tensile circumferential (hoop) stress $\tilde{\sigma}$. Given the curved nature of these streamlines, this stress leads to inward-directed radial forces on the network. 
By symmetry, hoop stress and, more generally, diagonal elastic contributions to the stress tensor must be even in the applied strain, since they are independent of the direction of rotation of the rheometer. Thus, to lowest order, a quadratic dependence on shear strain $\gamma$ is expected. We define this force (per unit volume) to be
\be
f_r=-\frac{\tilde\sigma}{r}\simeq-\frac{1}{r}\tilde AG\gamma^2,\label{hoop}
\ee
where $G$ is the shear modulus \sn{and} the coefficient $\tilde A>0$ is dimensionless. 
The minus sign and the inverse dependence on the radius $r$ account for the direction, as well as the curvature dependence.  
In an incompressible medium, in which net radial motion is not possible, this radial force must be balanced by a pressure that builds up toward a maximum along the axis of rotation. In the case of a free surface, as opposed to a rheometer plate, this gives rise to the well-known rod-climbing behavior \cite{Barnes1989}. In the case of a rheometer, the pressure results in a positive, compressive thrust $F$ in the axial direction. By contrast, if the network is compressible, as for multicomponent systems, then such stresses may relax by inward displacement of the network, as sketched in Fig.\ \ref{fig:1}.

In order to model the relaxation of hoop stress in a hydrogel, we use the minimal two-fluid model \cite{Brochard1977,Milner1993,Gittes1997,Levine2000}. Considering their biphasic nature, both synthetic hydrogels and biopolymer gels can be represented by this phenomenological model, in which the network displacement $\vec u$ and solvent velocity $\vec v$ are viscously coupled. 

The equation for the net force per unit volume acting on the fluid in the non-inertial limit is 
\be\label{eq:fluid_eq}
0=\eta\nabla^2\vec v-\vec\nabla P-\Gamma\left(\vec v-\dot{\vec{u}}\right),
\ee
where $\eta$ is the solvent viscosity and $P$ is the pressure. 
The corresponding equation for the net force on the network is
\be\label{eq:net_eq}
0=G\nabla^2\vec u+(G+\lambda)\vec\nabla\cdot(\vec\nabla\cdot\vec u)+\Gamma\left(\vec v-\dot{\vec{u}}\right),
\ee
where the shear modulus $G$ and Lam\'e coefficient $\lambda$ are assumed to be of the same order.
The viscous coupling constant $\Gamma$ is expected to be of order $\eta/\xi^2$ for a network with mesh or pore size $\xi$. \sn{This can be estimated by considering the drag on a total length $\sim\xi$ of polymer in a volume $\sim\xi^3$ moving with relative velocity $\vec v-\dot{\vec{u}}$ in a free-draining approximation.}

If the volume fraction of the network is small, as it is for most biopolymer gels ($\sim10^{-3}$), then to a good approximation \sn{the moving network displaces a negligible volume of fluid and the fluid phase remains incompressible, with $\vec\nabla\cdot\vec v=0$. Thus,} we can assume that the hoop stress $\tilde\sigma$ drives the network to move radially, against a solvent that is stationary in the radial direction. 
In Fig.\ \ref{fig:1}, $\vec u$ (red inward arrows) shows an inward, radial contraction of the network relative to the solvent upon shearing the gel placed between the gap of the cone-plate rheometer. In this case, \sn{with no-slip boundary conditions on the network, the radial component $u_r$ of the network displacement gives rise to stress and a restoring force that can be estimated from Eq.\ \eqref{eq:net_eq}. For a rheometer gap of thickness $d\ll r$, such as in a cone-plane rheometer, the axial gradients in Eq.\ \eqref{eq:net_eq} should be dominant, leading to an elastic contribution to the restoring force of order $Gu_r/d^2$. 
Thus, for a cone-plate rheometer with small cone angle $\alpha$, in which $d=r\tan\alpha\ll r$, the 
restoring force can be estimated as ${K}u_r/{r^2}$, where the phenomenological 
coefficient 
\be
K\simeq \frac{G}{\tan\left(\alpha\right)^2}.
\ee 
Together with the viscous drag on the network moving relative to a solvent that is stationary in the radial direction, the net radial component of the force per unit volume acting on the network can be written as}
\be\label{eq:simpnet_eq}
0=-\frac{K}{r^2}u_r-\frac{1}{r}\tilde\sigma-\Gamma\dot u_r,
\ee
which combines with Eqs.\ \eqref{hoop} and \eqref{eq:fluid_eq} to give
\be
\nabla_rP=\Gamma\dot u_r=-\frac{K}{r^2}u_r-\frac{1}{r}\tilde AG\gamma^2,\label{2Fluid}
\ee
where the strain $\gamma$ is independent of $r$ for a cone-plate geometry.
\sn{Corrections to Eq.\ \eqref{2Fluid}, from both $\nabla^2\vec u$ and $\vec\nabla\cdot(\vec\nabla\cdot\vec u)$ terms in Eq.\ \eqref{eq:net_eq}, are smaller by of order ${d}/{r^2}$.} 
The characteristic relaxation time implicit in Eq.\ \eqref{2Fluid} is then 
\be
\tau\sim \frac{\eta d^2}{(G\xi^2)}.\label{tau}
\ee
\sn{For a cone-plate rheometer such as we use here, this suggests a non single-exponential relaxation, since $d$ varies with $r$, resulting in a range of relaxation times (see section IIB). For a parallel-plate rheometer, a single-exponential relaxation may be expected. However, since the strain in this case is not uniform in $r$, the force in Eq.\ \eqref{hoop} will cease to vary as $1/r$.}

\subsection{Incompressible or strong coupling limit}
First, we consider the case of an incompressible medium, corresponding to the limit of strong coupling $\Gamma\rightarrow\infty$ and $u_r\rightarrow0$. Here, the network effectively inherits the incompressibility of the solvent and 
\be
\nabla_rP=-\frac{1}{r}\tilde\sigma.\label{phenomenology}
\ee
This pressure gradient will lead to a positive normal stress (thrust) contribution measured by the rheometer. 
Eq.\ \eqref{phenomenology} can be integrated to give
\be
P(R)-P(r)=-\tilde\sigma\log\left(\frac{R}{r}\right),
\ee
where $P(R)$ is the pressure at the sample boundary, i.e., atmospheric pressure $P_0$. 
The excess pressure,
\be
\Delta P=P(r)-P_0
\ee
can be integrated to give a positive (upward) contribution to the thrust $F$ 
\be
\int_0^R 2\pi r \Delta P\; dr=2\pi\tilde\sigma\int_0^R r\log\left(\frac{R}{r}\right)dr=\frac{\pi R^2}{2}\tilde\sigma.\label{PThrust}
\ee
Adding this to the direct contribution
\be
-\pi R^2\sigma_{zz}\label{zThrust}
\ee
from
$\sigma_{zz}$, we find that the normal stress, as reported by a cone-plate rheometer 
\be
\sigma_N\equiv\frac{2F}{\pi R^2}\label{sigmaNdef}
\ee
is given by
\be
\sigma_N=N_1=\sigma_{xx}-\sigma_{zz}\simeq \left(A_x-A_z\right)G\gamma^2,\label{N1incomp}
\ee
implying that
\be
\tilde\sigma=\sigma_{xx}+\sigma_{zz}\simeq \tilde AG\gamma^2,
\ee 
where $\tilde A=\left(A_x+A_z\right)$. 
In Eq.\ \eqref{N1incomp} we have assumed not only incompressibility of the medium, but also the standard relationship between the thrust $F$ and the first normal stress difference $N_1\equiv\sigma_{xx}-\sigma_{zz}$ \cite{Venerus2007}, valid for incompressible materials and a cone-plate rheometer. We have used this assumption to identify $\tilde\sigma$ in Eq.\ \eqref{phenomenology}. 
Although this relationship between $F$ and $N_1$ is a standard result for the cone-plate geometry, it is worth noting that this can change, depending on the shape of the sample/air interface, or with finite surface tension \cite{Venerus2007}. 
In the next section, we also show how this relationship can be violated for compressible networks, such as hydrogels. 
Nevertheless, because this relationship is so standard in rheology, with rheometers usually reporting the thrust $F$ as $N_1$, we will use Eq.\ \eqref{sigmaNdef} to express the normal stress in the following sections. Importantly, however, for multi-component systems such as hydrogels, this should be considered an effective or apparent $N_1$, i.e., as reported by a rheometer, which may or may not be equal to the actual stress difference $N_1=\sigma_{xx}-\sigma_{zz}$.

As noted, the various normal stress components are expected to have leading $\gamma^2$ behavior, while the shear modulus $\sigma_{xz}\simeq G\gamma$ in the linear (shear) elastic regime. Thus, we define
\be
\sigma_{xx}\equiv A_xG\gamma^2\quad\mbox{and}\quad\sigma_{zz}\equiv A_zG\gamma^2
\ee
\sn{Usually, $\sigma_{xx}$ is of order but larger than $\sigma_{zz}$ in magnitude, due to the increasing alignment of fibers into the shear direction with increasing strain \cite{Larson1998}.  As defined, both stress components are strictly positive (tensile). 
Thus, we expect that $A_x\gtrsim A_z$ and $N_1>0$ \cite{Lodge1972}. 
For semiflexible gels, we expect $A_x\gtrsim A_{z}\sim {1}/{\gamma_0}$, based on the prior low-frequency model \cite{Janmey2007,Kang2009}, where $\gamma_0$ represents the onset strain for nonlinear elasticity, which is typically of order 10\% for biopolymer networks.
These expectations, however, are based on the assumption of affine deformation, which may not be valid for some stiff polymer gels \cite{heussinger2007nonaffine,Broedersz2014}.}

\subsection{Compressible limit of hydrogels}
\begin{figure}[htbp]
\centering
\includegraphics[width=8cm]{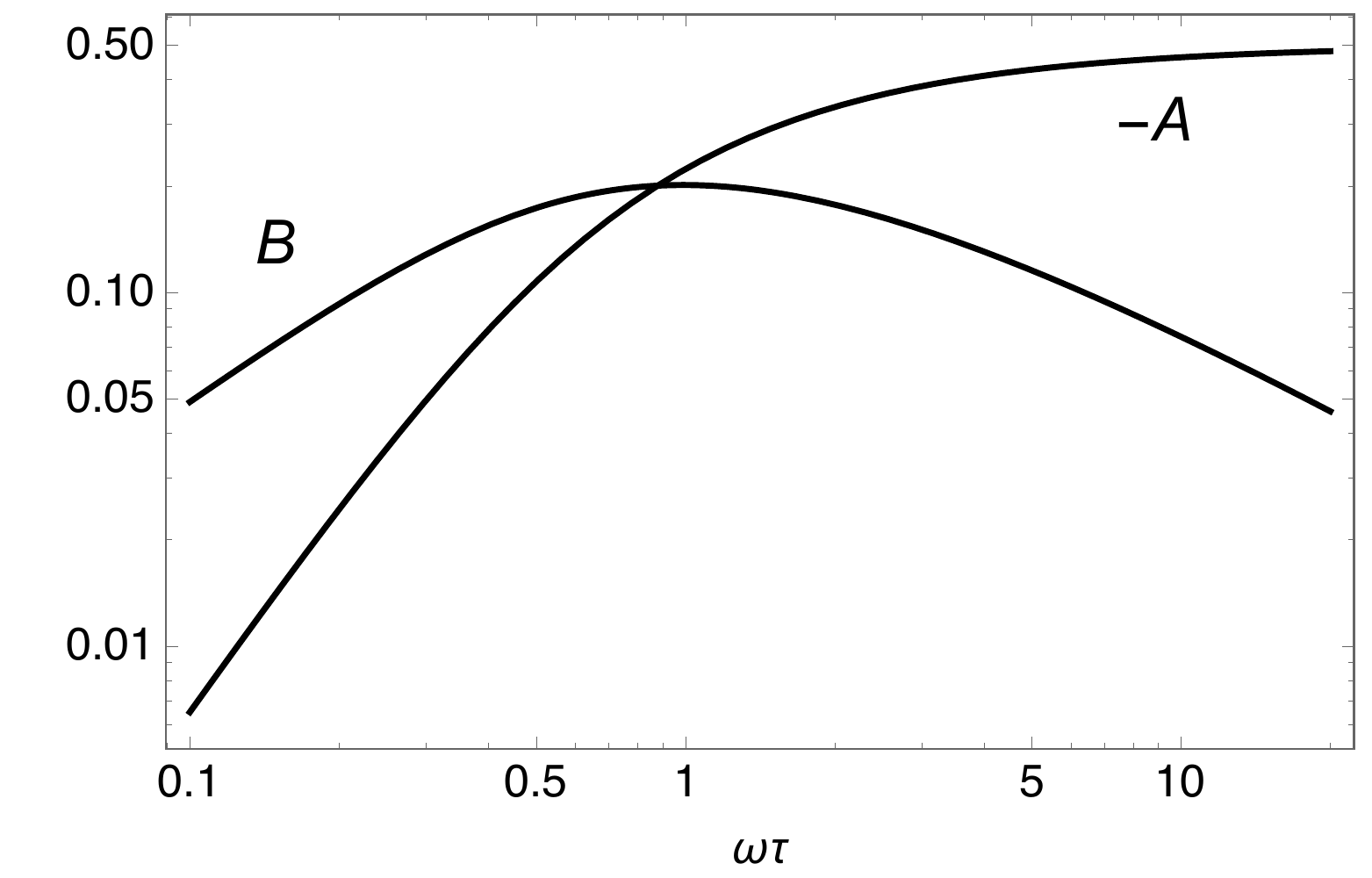}
\caption{Plot of $-\mathcal{A}$ from Eq.\ (\ref{calA}) and $\mathcal{B}$ from Eq.\ (\ref{calB}) vs $\omega\tau$.}
\label{fig:NormStress1}
\end{figure}

In the limit of long times $t\gg\tau$ in Eq.\ \eqref{tau} and low frequencies $\omega\tau\ll 1$, $\dot u_r\rightarrow0$ in Eq.\ (\ref{2Fluid}). Here, $\Delta P$ vanishes and the apparent $N_1$ measured is that of
Refs.\ \cite{Janmey2007,Kang2009}
\be
\frac{2F}{\pi R^2}=\sigma_N%=N_1^{\mbox{\scriptsize app}}
=-2\sigma_{zz}=-2A_zG\gamma^2.\label{N1comp}
\ee
For intermediate times/frequencies, we solve Eq.\ (\ref{2Fluid}) for $u_r(t)$, with $\gamma(t)=\tilde\gamma\sin(\omega t)$. The net elastic force per volume on a network element must be balanced by its drag through the solvent, which sets up a pressure gradient in the solvent. Importantly, in spite of the nonlinear dependence on strain, Eq.\ \eqref{N1comp} remains a linear equation in $u_r$, albeit inhomogeneous. The long-time, intermediate frequency steady state (ss) solution to this is given by
\begin{widetext}
\be
u_r^{\mbox{\scriptsize (ss)}}(t)
=-\frac{\tilde AG_0\tilde\gamma^2 r \left(-K^2 \cos (2 t \omega )+K^2-2 \Gamma  K r^2 \omega  \sin
   (2 t \omega )+4 \Gamma ^2 r^4 \omega ^2\right)}{2 \left(K^3+4 \Gamma ^2
   K r^4 \omega ^2\right)}.
\ee\vspace{2mm}
Using this and $\nabla_rP=\Gamma\dot u_r$ we find
\be
P^{\mbox{\scriptsize (ss)}}(r)=
\frac{\tilde AG_0\tilde\gamma^2}{8} A \left(\cos (2 t \omega ) \log \left(K^2+4 \Gamma ^2 r^4 \omega ^2\right)-2
   \sin (2 t \omega ) \tan ^{-1}\left[\frac{2 \Gamma  r^2 \omega }{K}\right]\right)+g(t),\label{Pss}
\ee
where $g(t)$ is a constant of integration with respect to $r$, although a function of $t$, which is determined by $P(R)=P_0$ as above. 
After a further integration of $\Delta P=P(r)-P_0$, as in Eq.\ \eqref{PThrust}, and combining with Eq.\ \eqref{zThrust}, we find the steady-state
\be
\sigma_N^{\mbox{\scriptsize (ss)}}=-2A_zG\tilde\gamma^2\sin^2(\omega t)+\tilde AG\tilde\gamma^2 \left({\mathcal A}\cos (2 \omega t)+{\mathcal B}\sin (2\omega t)\right),\label{N1app}
\ee
where
\be
{\mathcal A}=-\frac{1}{8 \omega\tau  }
\Bigg[2 \tan ^{-1}\left(1+2 \sqrt{\omega\tau}\right)
   +2 \tan ^{-1}\left(1-2 \sqrt{\omega\tau}\right)
   -\pi +4 \omega\tau\Bigg]\label{calA}
   \ee
   and
\end{widetext}
\be
{\mathcal B}=\frac{1}{8\omega\tau }\log\left(1+4\omega^2\tau^2\right).\label{calB}
\ee
Figure \ref{fig:NormStress1} shows the parameters $-\mathcal A$ and $\mathcal B$ versus $\omega\tau$. Both of these dimensionless coefficients vanish in the low frequency or fully compressible limit, leaving only the first (axial stress) term on the right hand side of Eq.\ \eqref{N1app}.

In addition to the steady-state solution for $u_r(t)$, there is also a transient contribution $u_r^{\mbox{\scriptsize (tr)}}(t)$, which can be found by choosing a homogeneous solution of Eq.\ \eqref{2Fluid} such that $u_r(t)=u^{\mbox{\scriptsize (ss)}}(t)+u^{\mbox{\scriptsize (tr)}}(t)=0$ at $t=0$:
\be
u^{\mbox{\scriptsize (tr)}}(t)=
\frac{\tilde AG_0\tilde\gamma^2 r}{2 K \left(1+\frac{K^2}{4 \Gamma ^2 r^4 \omega ^2}\right)}e^{-\frac{K}{\Gamma  r^2}t}.\label{utrans}
\ee
This transient is most relevant to the case where its characteristic relaxation time $\tau\sim{\frac{\Gamma  R^2}{K}}$ is large compared with the period of oscillation $\sim\frac{1}{\omega}$. Thus, we neglect the second term in the denominator of Eq.\ \eqref{utrans} to find
\be
u^{\mbox{\scriptsize (tr)}}(t)\simeq\frac{\tilde AG_0\tilde\gamma^2 r}{2 K}e^{-\frac{K}{\Gamma  r^2}t},
\ee
from which we determine
\be
\nabla_rP^{\mbox{\scriptsize (tr)}}=\Gamma\dot u_r\simeq-\frac{\tilde AG_0\tilde\gamma^2}{2 r}e^{-\frac{K}{\Gamma  r^2}t}\label{gradPtransient}
\ee
and the transient contribution to $\sigma_N$
\bea
\sigma_N^{\mbox{\scriptsize (tr)}}&\simeq&\frac{1}{2} \tilde AG_0\tilde\gamma^2 \left( \frac{t}{\tau}\text{Ei}\left(-\frac{t}{\tau}\right)+e^{-\frac{t}{\tau}}\right)\nonumber \\
&\simeq&\frac{1}{2} \tilde AG_0\tilde\gamma^2\exp\left[{-1.91\left(\frac{t}{\tau}\right)^{0.78}}\right].\label{N1trans}
\eea
where $\text{Ei}\left(x\right)$ is the exponential integral function. As can be seen either from $\tau\sim \frac{\eta d^2}{(G\xi^2)}$, which depends on the gap $d$, or from the time dependence of Eq.\ \eqref{gradPtransient}, there is no single relaxation time. Thus, $\sigma_N^{\mbox{\scriptsize (tr)}}$ can be well approximated by a stretched exponential response. The final approximation in Eq.\ \eqref{N1trans} is valid to within less than 2\% until the transient has decayed to less than 2\% of its initial value. 

The general expression for $\sigma_N$ is given by the sum of Eqs.\ \eqref{N1app} and \eqref{N1trans}.
For an incompressible system, Eq.\ \eqref{N1incomp} is recovered for $u_r(t)=u^{\mbox{\scriptsize (ss)}}(t)+u^{\mbox{\scriptsize (tr)}}(t)$ as $\tau\rightarrow\infty$.
In the limit of low frequency and long times, the steady-state solution reduces to the fully compressible limit of Eq.\ \eqref{N1comp}. 

\section{Calculation of stress components for semiflexible gels}\label{C}
In the model above, the thrust F measured on the top plate of the rheometer depends on the leading-order, $\sim\gamma^2$ dependence of $\sigma_{xx}$ and $\sigma_{zz}$. We calculate these for semiflexible polymer networks within the affine approximation, in which the stress arises from the longitudinal compliance of polymer segments. For an inextensible chain, the entropic response comes from the thermal bending fluctuations of the filament \cite{MacKintosh1995,Gardel2004,Storm2005,Broedersz2014}. It is often useful to consider the limit of large persistence length $\ell_p$ that is much larger than the length of segments between cross-links, $\ell$, which we assume to be constant. 
Perhaps surprisingly, even for $\ell\ll\ell_p$, the longitudinal response can be dominated by the transverse thermal fluctuations. In this limit, the filament is nearly straight, with only small transverse fluctuations \cite{Head2003}. Reconstituted fibrin networks constitute a prominent example, having typical persistence lengths of tens of $\mu m$, much larger than the $\mu m$ scale cross-link distance \cite{Piechocka2016}.

In the presence of longitudinal tension $f$ acting on a segment of length $\ell$, the thermal average contraction of the segment is given by \cite{MacKintosh1995}
\begin{equation}
\langle\Delta\ell\rangle=\frac{kT\ell^2}{\kappa\pi^2}
\sum^\infty_{n=1}\frac{1}{n^2+\phi},\label{eq:contract}
\end{equation}
where $\kappa=kT\ell_p$ is the bending rigidity and  
\begin{equation}
\phi=\frac{f\ell^2}{(\kappa\pi^2)}
\end{equation}
is a dimensionless measure of force. 
In the absence of tension $f$, the contraction reduces to $\langle\Delta\ell\rangle_0=\frac{{\ell^2}}{({6\ell_p})}$, which also represents the full extension/compliance in the limit of high force. 
This scaling can be anticipated, by noting that the mean-square transverse fluctuations $\langle u_\perp^2\rangle$ should be proportional to $kT$ and inversely proportional to $\kappa$. Thus, $\langle u_\perp^2\rangle\sim\ell^2/\ell_p$. These longitudinal fluctuations give rise to $ \langle\Delta\ell\rangle_0$, thus we expect $\langle\Delta\ell\rangle_0\sim\ell/\ell_p$. 

For a finite longitudinal tension $f$, the extension of the chain segment (toward full extension $\langle\Delta\ell\rangle_0$) is given by 
\begin{equation}
\delta\ell(f)=\langle\Delta\ell\rangle_0-\langle\Delta\ell\rangle
=\frac{\ell^2}{\pi^2\ell_p}\sum_n\frac{\phi}{n^2\left(n^2+\phi\right)}.\label{eq:extension1}
\end{equation}
The sum above can be evaluated, to give
\begin{equation}
\delta\ell(f)
=\frac{\ell^2}{\pi^2\ell_p}\Gamma(\phi),\label{eq:extension2}
\end{equation}
where
\be
\Gamma(\phi)=\frac{\pi ^2 \phi -3 \pi  \sqrt{\phi } \coth \left(\pi  \sqrt{\phi
   }\right)+3}{6 \phi }.
\ee
From the inverse function $\Gamma^{-1}$, the force-extension curve is 
\be
f\left(\delta\ell\right)=\frac{\kappa\pi^2}{\ell^2}\phi=\frac{\kappa\pi^2}{\ell^2}\Gamma^{-1}\left(\frac{\pi^2\ell_p}{\ell^2}\delta\ell\right).
\ee
In practice, this inversion needs to be done numerically. One can, however, determine this term by term in an expansion about $\delta\ell=0$, e.g., as
\be f=\mu_1\frac{\delta\ell}{\ell}+\mu_2\left(\frac{\delta\ell}{\ell}\right)^2\cdots,\ee
where the 1D Young's modulus \cite{MacKintosh1995}
\be
\mu_1=\frac{\kappa\ell_p\pi^4}{\ell^3}\frac{1}{\Gamma'(0)}=\frac{90\kappa\ell_p}{\ell^3}
\ee
and
\be
\mu_2=\frac{\kappa\ell_p^2\pi^6}{\ell^4}\left(\frac{-\Gamma''(0)}{\left(\Gamma'(0)\right)^3}\right)=\frac{5400\kappa\ell_p^2}{7\ell^4}
\ee
Here, it is important to notice that the longitudinal strain $\delta\ell/\ell$ \sn{on each segment is bounded above by $\ell/(6\ell_p)$, since $\delta\ell<\langle\Delta\ell\rangle_0$.
Thus, in the semiflexible limit $\ell\ll\ell_p$, nonlinearities are expected to appear at small strains of order $\ell/\ell_p$. 

Using a variant of the Kirkwood formula for the stress, $\sigma$, in terms of multiple segments \cite{Larson1998}
\be
\sigma_{ij}=\frac{1}{V}\sum_\beta r^{(\beta)}_if^{(\beta)}_j,
\ee
where $V$ is the sample volume and the sum is over all segments $\beta$. The segment lengths are $|\vec r^{\;(\beta)}|$ and the orientations are $\hat r^{(\beta)}$. The (tensile) force in segment $\beta$ is $\vec f^{\;(\beta)}$. Since this force $\vec f^{\;(\beta)}=\hat r^{(\beta)}f^{(\beta)}$ is directed along the segment, the stress can be expressed as \cite{Gittes1998,Morse1998}
\be
\sigma_{ij}=\rho\langle f n_in_j\rangle,\label{Kirkwood}
\ee
where $\rho$ is the total length of polymer per unit volume and $\langle\cdots\rangle$ represents an average over all segment orientations, which we represent as $\vec n=\left(\sin(\theta)\cos(\phi),\cos(\theta)\sin(\phi),\cos(\theta)\right)$ in terms of the usual polar and azimuthal angles. 

Simple volume preserving shear strain $\gamma$ in the $x$ direction with gradient in the $z$ direction can be represented by the deformation gradient tensor
\be
\Lambda=\left(\ba{ccc}
1&0&\gamma\\
0&1&0\\
0&0&1\ea\right).
\ee
To linear order in the strain, the relative segment extension $\delta\ell/\ell=\sin(\theta)\cos(\theta)\cos(\phi)\gamma$. 
To this order, $f=\mu_1\sin(\theta)\cos(\theta)\cos(\phi)\gamma$  in Eq.\ \eqref{Kirkwood} and}
\be
\sigma_{xz}=\rho\mu_1\gamma\left\langle\cos(\theta)^2\sin(\theta)^2\cos(\phi)^2\right\rangle=\frac{1}{15}\rho\mu_1\gamma.\label{shearstress}
\ee
\noindent By symmetry, corrections to this will only involve odd powers of strain $\gamma$. 

The various normal stress components can be calculated similarly \cite{Janmey2007,Kang2009}, e.g., with
\be
\sigma_{zz}=\frac{\rho\kappa\pi^2}{\ell^2}\left\langle\cos(\theta)^2\Gamma^{-1}\left(\frac{\pi^2\ell_p}{\ell}\sin(\theta)\cos(\theta)\cos(\phi)\gamma\right)\right\rangle.
\ee
where, by symmetry only even terms in the expansion of $\Gamma^{-1}$ can contribute. Thus, the general form of the non-zero terms involves the average
\be
\langle n_z^2\left[n_xn_z\right]^{2m}\rangle=\left\langle\cos(\theta)^2\left[\sin(\theta)\cos(\theta)\cos(\phi)\gamma\right]^{2m}\right\rangle,
\ee
where $m=1, 2, 3, \ldots$.
The corresponding averages
\be
\langle n_x^2\left[n_xn_z\right]^{2m}\rangle
\ee
contributing to $\sigma_{xx}$ are identical by symmetry. 
Thus, we can see that, in the extreme semiflexible limit $\ell\ll\ell_p$
\be\sigma_{xx}=\sigma_{zz}.\ee
The lowest-order contribution to these is 
\bea
\sigma_{xx}&=&\sigma_{zz}=\rho\mu_2\gamma^2\left\langle\cos(\theta)^2\left[\sin(\theta)\cos(\theta)\cos(\phi)\right]^{2}\right\rangle\nonumber\\
&=&\frac{1}{35}\rho\mu_2\gamma^2\\
&=&\frac{180\ell_p}{49\ell}G\gamma^2\nonumber,
\eea 
where $G=\rho\mu_1/15$. 

\begin{figure}[htbp]
\centering
%\vspace{.1cm}
\includegraphics[width=8cm]{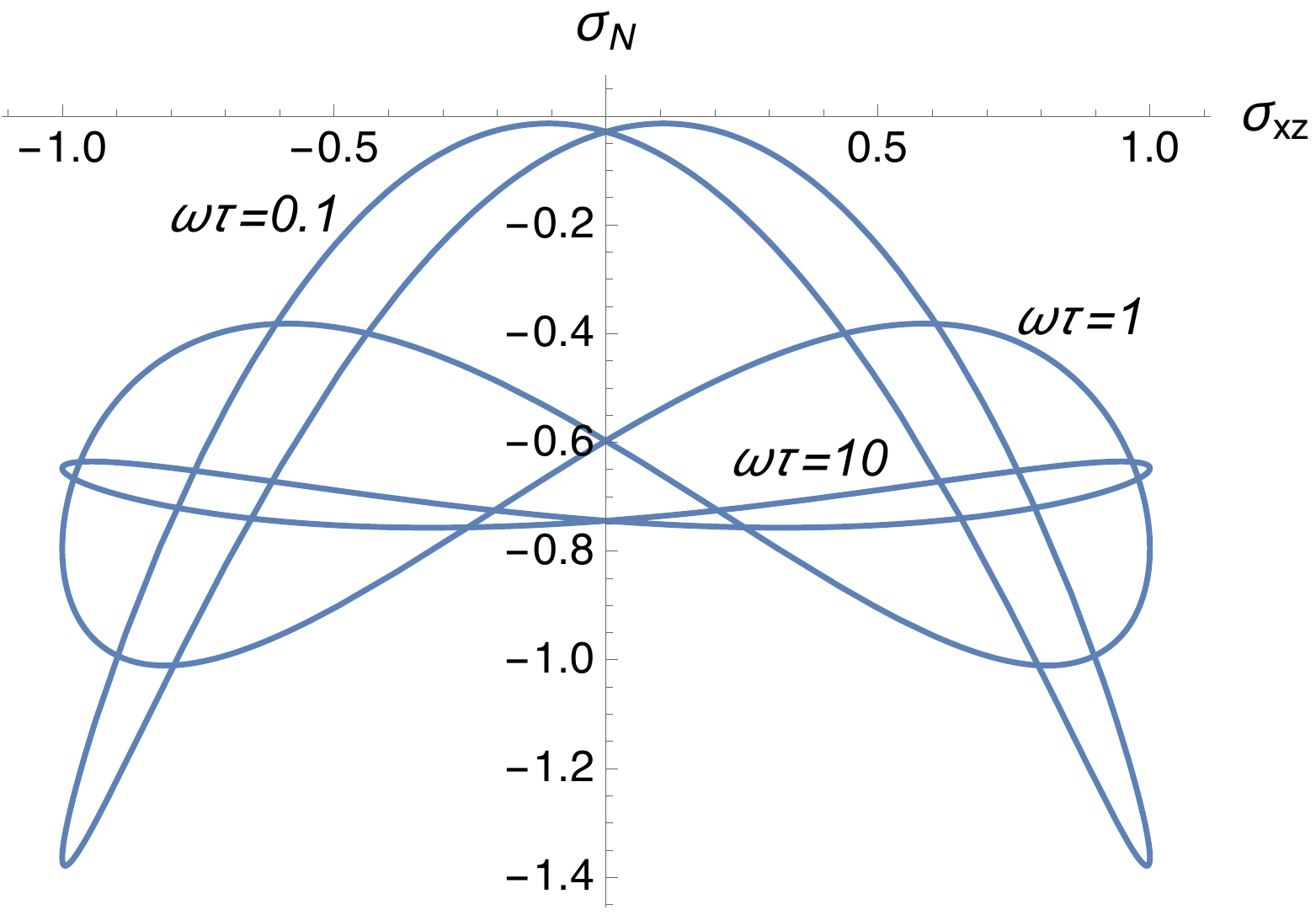}
\caption{A series of Lissajous figures for $\omega\tau=$0.1, 1, 10. The shear stress in Eq.\ \eqref{shearstress} 
has been normalized by its maximum, $G\tilde\gamma$, and the steady-state $\sigma_N$ in Eq.\ \eqref{N1app} has been normalized by $G\tilde\gamma^2/\gamma_0$. Here, we have also used Eqs.\ \eqref{Az_gamma0} and \eqref{1param} to calculate $A_z$ and $\tilde A$ for $\gamma_0=0.1$}
\label{fig:NormStress3}
\end{figure}

In addition to the contributions to $\sigma_{xx}$ and $\sigma_{zz}$ above, which come from the intrinsically nonlinear stretching response of semiflexible chains and are dominant in the limit of $\ell\ll\ell_p$, there are additional terms arising from purely geometric nonlinearities \cite{Morse1999,Storm2005,Cioroianu2013}. 
\sn{For the deformation gradient above, the relative extension $\delta\ell/\ell$ is determined from the deformed $\vec n'$ according to}
\bea\delta\ell/\ell&=&|\vec n'|-1\\
&=&\gamma  \sin (\theta ) \cos (\theta ) \cos
   (\phi )\nonumber\\
   &&+\frac{1}{2} \gamma ^2 \left(\sin ^2(\theta ) \cos ^2(\theta ) \sin ^2(\phi
   )+\cos ^4(\theta )\right)\cdots.\nonumber
\eea
From this, \cite{Morse1999,Storm2005,Cioroianu2013}
\bea
\sigma_{xx}&=&\rho\left\langle \frac{n'_xn'_x}{|\vec n'|}\frac{\kappa\pi^2}{\ell^2}\Gamma^{-1}\left(\frac{\pi^2\ell_p}{\ell}\left(|\vec n'|-1\right)\right)\right\rangle\nonumber\\
&=&\frac{1}{105} \rho\gamma^2 (13 \mu_1+3 \mu_2)\\
&=&\left(\frac{180\ell_p}{49\ell}+\frac{195}{105}\right)G\gamma^2,\nonumber
\eea
\bea
\sigma_{yy}&=&\rho\left\langle \frac{n'_yn'_y}{|\vec n'|}\frac{\kappa\pi^2}{\ell^2}\Gamma^{-1}\left(\frac{\pi^2\ell_p}{\ell}\left(|\vec n'|-1\right)\right)\right\rangle\nonumber\\
&=&\frac{1}{105} \rho\gamma^2 (2 \mu_1+2 \mu_2)\\
&=&\left(\frac{120\ell_p}{49\ell}+\frac{30}{105}\right)G\gamma^2,\nonumber
\eea
and
\bea
\sigma_{zz}&=&\rho\left\langle \frac{n'_zn'_z}{|\vec n'|}\frac{\kappa\pi^2}{\ell^2}\Gamma^{-1}\left(\frac{\pi^2\ell_p}{\ell}\left(|\vec n'|-1\right)\right)\right\rangle\nonumber\\
&=&\frac{1}{35} \rho\gamma^2 (2 \mu_1+ \mu_2)\\
&=&\left(\frac{180\ell_p}{49\ell}+\frac{30}{35}\right)G\gamma^2.\nonumber
\eea
I.e.,
\be A_z=\left(\frac{180\ell_p}{49\ell}+\frac{30}{35}\right)%=\left(\frac{60}{49\gamma_{\mbox{\scriptsize max}}}+\frac{30}{35}\right)\nonumber\\
\simeq\left(\frac{0.61}{\gamma_0}+0.86\right) \label{Az_gamma0}\ee
and
\be\tilde A=2A_z+1 \label{1param}\ee
above, where $\gamma_0$ is the strain at the onset on nonlinearity, defined as the point at which $d\sigma_{xz}/d\gamma$ increases by a factor of $\simeq 2$ above its linear value, $G$.
Figure \ref{fig:NormStress3} shows the steady-state $\sigma_N^{\mbox{\scriptsize (ss)}}$ in Eq.\ \eqref{N1app} versus shear stress for various values of $\omega\tau$, where we have used the specific predictions in Eqs.\ \eqref{Az_gamma0} and \eqref{1param} for $\gamma_0=0.1$.

Interestingly, we find that the Lodge-Meissner relation 
\begin{equation}
N_1=\sigma_{xz}\gamma
\end{equation}
still holds for the various stress terms calculated above 
in the incompressible limit \cite{Lodge1972}. 
Although this relation is derived for flexible polymer systems, it is expected to be valid even for the present model of semiflexible polymers, since this model assumes both purely central force (polymer stretching) response and purely affine deformation. 

\section{Experimental}\label{Exp}

Fibrin gels were polymerized from human plasma fibrinogen and $\alpha$-thrombin. Fibrin was polymerized in a buffer containing 150 mM~NaCl, 20~mM HEPES and 5~mM CaCl$_2$, at pH 7.4. Fibrinogen stock solution was diluted in the assembly buffer to reach a final concentration of 8~mg/mL. Polymerization was initiated by the addition of 0.5~U/mL thrombin. The samples were then transferred to the rheometer geometry where the polymerization reaction occurred at a specified temperature (22\textdegree C, 27\textdegree C or 37\textdegree C) for at least 12 hours. All chemicals were bought from Sigma Aldrich (Zwijndrecht, The Netherlands); fibrinogen and thrombin were purchased from Enzyme Research Laboratories (Swansea, United Kingdom).

We used an MCR 302 rheometer (Anton Paar, Graz, Austria) with stainless steel cone-plate geometry (40~mm diameter, 2\textdegree) for all normal force measurements. A solvent trap was used to prevent evaporation during the measurement, in addition to a small layer of low viscosity mineral oil added around the sample. During polymerization, a small shear oscillation (amplitude 0.1\% and frequency 1~Hz) was applied to monitor the evolution of the storage modulus. The unprocessed, time-dependent normal force response to an applied shear was recorded using an oscilloscope Tectronix DPO 3014 plugged to the analogue outputs of the rheometer. The applied stress was 800~Pa and the shearing frequency varied from 0.001~Hz to 7~Hz. To obtain the differential modulus $K'$ as a function of applied shear strain, we used a MCR 501 rheometer with a 40~mm, 1\textdegree~cone-plate geometry and applied a stepwise increasing shear stress with a superimposed oscillatory strain with amplitude 10\% of the constant shear level.

\section{Results and Discussion}\label{Results}
\begin{figure}[htbp]
	\includegraphics[width=0.87\columnwidth]{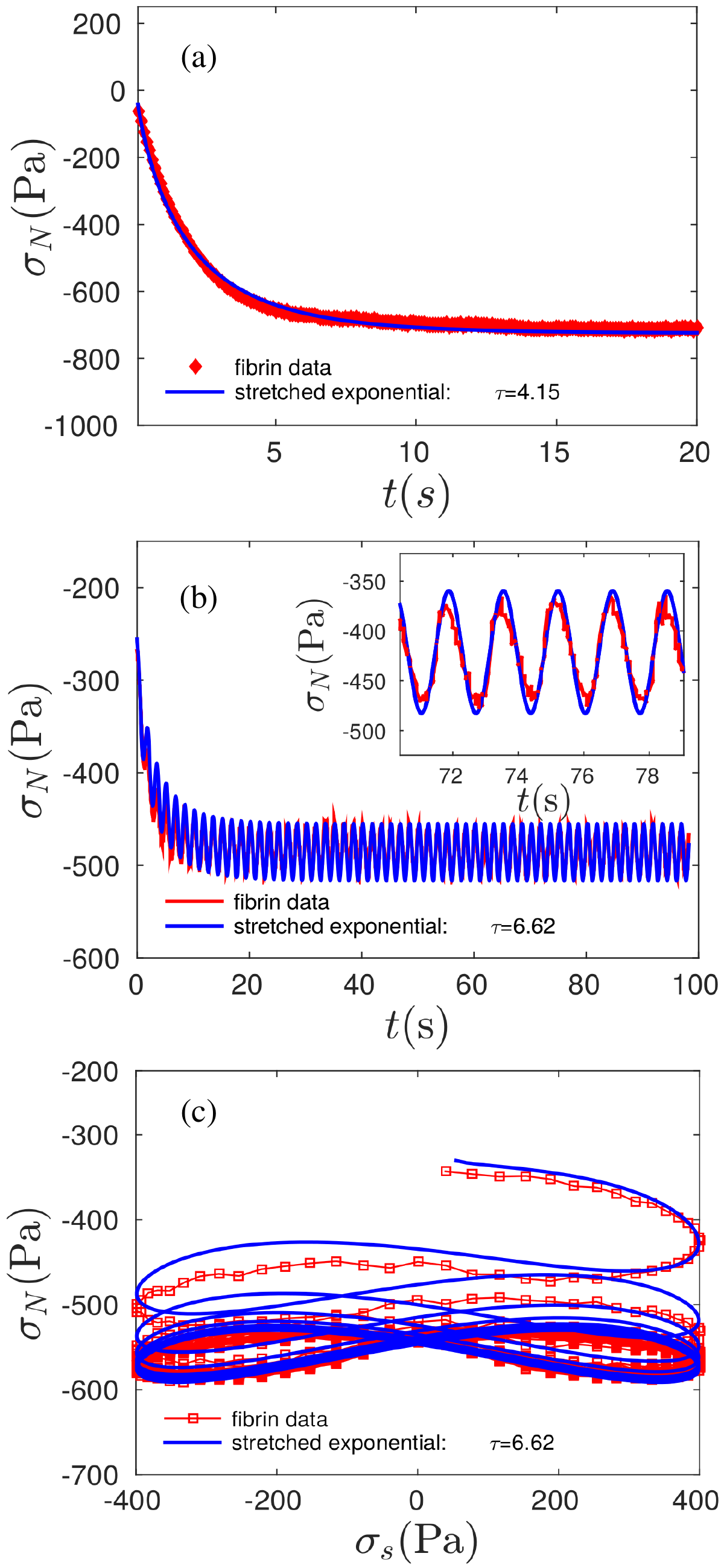}
	\caption{(color online) Fibrin gels polymerized at 22\textdegree C. (a) The red line shows the \sn{normal stress relaxation versus time for a constant shear stress. The blue line is a stretched exponential fit to the data, which} yields the relaxation time $\tau=4.15s$. (b) The red line shows the \sn{normal stress versus time for an oscillatory shear stress with frequency $\nu= 0.3 Hz$. The blue line is the fit using Eq.\ \eqref{N1trans}, which yields the relaxation time $\tau=6.62s$. The inset zooms in on the steady-state response. (c) The same data shown in (b) plotted as normal stress versus shear stress.}}
	\label{fig:5}
\end{figure}

The model presented above predicts a transient response in the normal stress at the beginning of the shearing process. 
We test this by measuring the full time dependence of $\sigma_N$, as determined by the thrust $F$, according to Eq.\ \eqref{sigmaNdef}, 
as shown in Fig.\ \ref{fig:5}. In Fig.\ \ref{fig:5}a, a constant shear stress is applied whereas in Fig. \ref{fig:5}b we show the transient normal stress response to an oscillatory shear stress. The red line shows the experimental data, the blue line is the fit using Eq.\ \eqref{N1trans}. As predicted by the theory, in both panels, the normal stress decays until it reaches a steady state, where the fitted decay constants are very comparable between the experiments with the constant shear stress and the oscillatory shear stress. In Fig.\ \ref{fig:5}c the data of Fig.\ \ref{fig:5}b are replotted, with $\sigma_N$ as a function of shear stress instead of time. This representation (also known as Lissajous curve) allows us to have a better perspective of the initial transient behavior found in both the experimental data and the model. 

\begin{figure*}[htbp]
	% Note that the * is needed to properly display the figure over the entire width of the page
	\includegraphics[width=1.9\columnwidth]{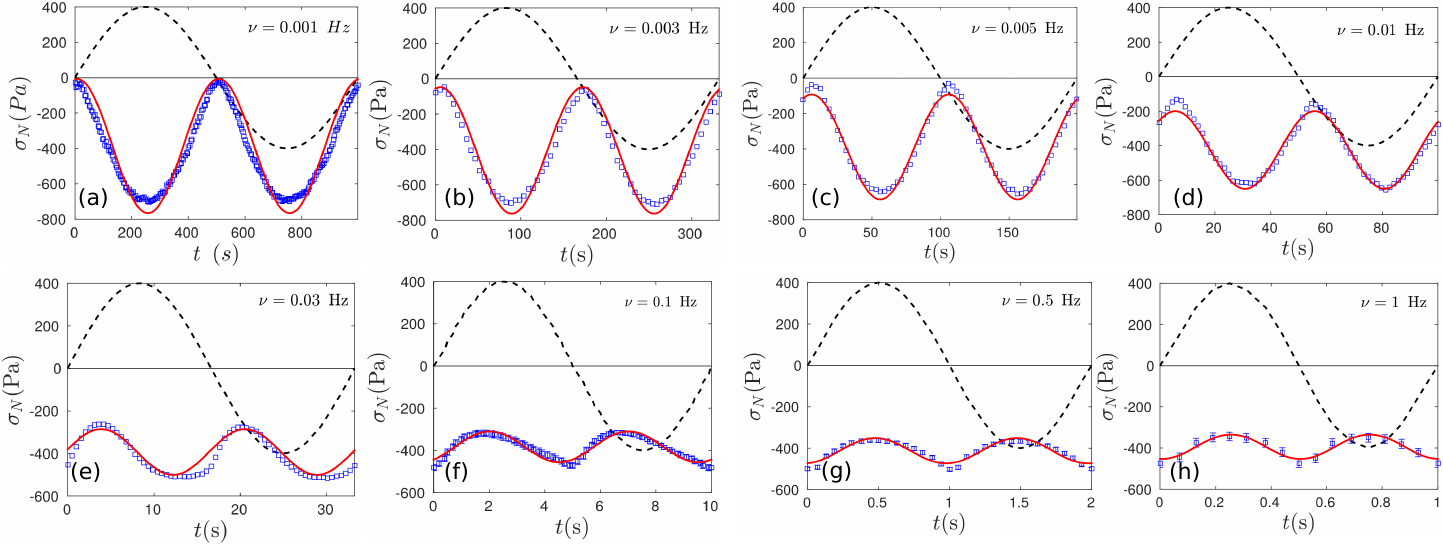}
	\caption{
		(color online) Normal stress response (blue square symbols) of a fibrin gel polymerized at 27\textdegree C to an oscillating shear stress (dashed line) for different frequencies $\nu$ versus time. The fit (red line) according to Eqs.\ (\ref{N1app}-\ref{calB}) is also shown where the fitting parameters are $2 A_z$ and ${\tilde{A}}$. In these equations the shear modulus $G$ is independently obtained from the rheology data. The data shown at $\nu>0.1$~Hz represent averages with standard deviations obtained by averaging over 34 cycles to compensate for the low sampling frequency of the rheometer. \sn{(Data in (a), (d) and (h) from Ref.\ \cite{DeCagny2016}.)}}
	\label{fig:6}
\end{figure*}

Changing the polymerization temperature of the fibrin gels can change the mesh size of the network and therefore influence the characteristic time constant $\tau$ according to Eq.\ \eqref{tau}.  From the results in \cite{DeCagny2016}, fibrin gels polymerized at 22\textdegree C and 27\textdegree C are expected to have time constants $\tau$ of around 5 and 12s, respectively. The latter is especially interesting, since the frequency ($1/\tau\sim$0.08Hz) associated with this characteristic time is in the middle of frequency range accessible with our set-up (approximately 0.001 to 1~Hz). This allows us to probe the behavior of the gels at frequencies above and below the characteristic frequency. The normal stress response to oscillatory shear at different frequencies is plotted for these gels in Fig.\ \ref{fig:6} (blue squares), together with the applied shear stress (black dashed line) and the corresponding fits of the steady state oscillatory normal stress (Eq.\ \eqref{N1app}, red line), as functions of time. Here, the shear modulus $G$ is measured independently from the shear stress at small strain. Hence, the only fitting parameters are $2A_z$ and $\tilde{A}$. We use a common relaxation time $\tau= 26.9s$ for all data sets at 27\textdegree C. This is obtained by first fitting the datasets with $\tau$ as fitting parameter (together with $2A_z$ and $\tilde{A}$). Then the average is calculated of the values of $\tau$ over the frequency range where $\tau$ shows sensitivity to the frequency. Finally, this average relaxation time is used as a constant when the datasets are fitted again (Fig.\ \ref{fig:6}). Qualitatively, when the frequency of the applied oscillatory shear stress increases, the amplitude of the normal stress signal decreases, together with the average value of the normal stress. This is in agreement with Fig.\ \ref{fig:NormStress3}, which shows the theoretical signals for different values of $\omega\tau$. The fits follow this trend.

\begin{figure*}[htbp]
	% Note that the * is needed to properly display the figure over the entire width of the page
	\includegraphics[width=1.9\columnwidth]{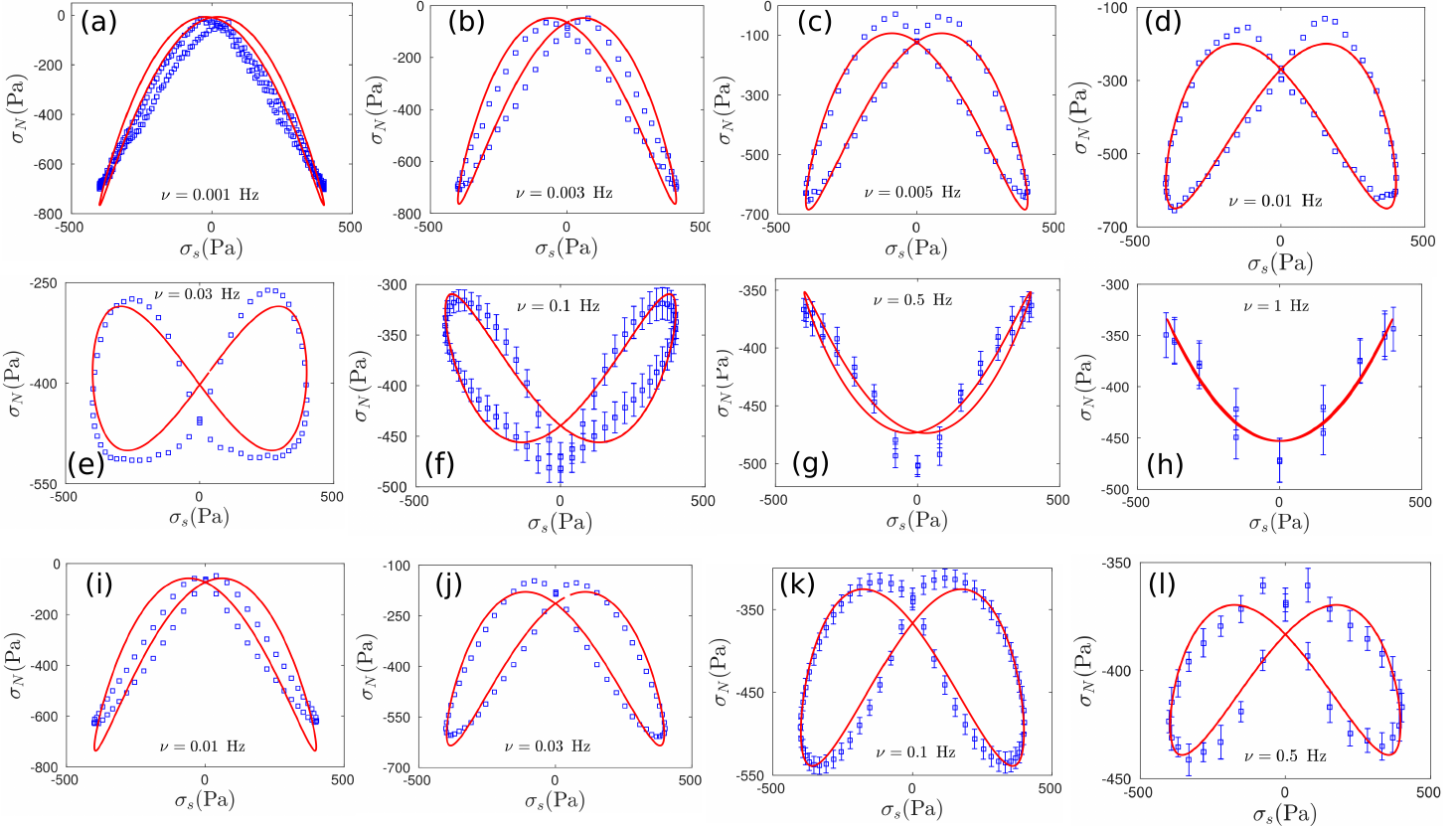}
	\caption{
		(color online) (a-h) Normal stress response (blue square symbols) of fibrin gels polymerized at 27\textdegree C for different shearing frequencies versus shear stress. The fit (red solid line) according to Eqs.\ (\ref{N1app}-\ref{calB}) is also shown. The data are the same as in Fig. \ref{fig:6}. \sn{(Data in (a), (d) and (h) from Ref.\ \cite{DeCagny2016}.)}
		(i-l) (color online) Normal stress response (blue square symbols) of a fibrin gels polymerized at 22\textdegree C for different shear stress frequencies versus shear stress. The fit (red solid line) according to Eqs.\ (\ref{N1app}-\ref{calB}) is also shown where the fitting parameters are $2 A_z$ and ${\tilde{A}}$. The characteristic time $\tau$ for this gel is shorter ($\tau=10.17$~s) than for the gel in (a), such that similar "butterfly-shapes" are found at a higher frequency. The data shown at $\nu>0.1$~Hz represent averages with standard deviations obtained by averaging over 34 cycles to compensate for the low sampling frequency of the rheometer.}
	\label{fig:7}
\end{figure*}

Perhaps the agreement between experimental data and the theoretical predictions plotted in Fig.\ \ref{fig:NormStress3} is easier to spot in Fig.\ \ref{fig:7}a where the normal stress response from Fig.\ \ref{fig:6} (blue square symbols) is plotted versus shear stress. As clearly seen in the figures, the Lissajous curves change shape as the frequency of the applied shear stress is increased. For oscillation periods longer than $\tau$ (low frequencies), the normal stress decreases with increasing shear stress, demonstrating contractile behavior under shear. In contrast, for oscillation periods shorter than $\tau$ (high frequencies), the normal stress increases with increasing shear stress, demonstrating extensile behavior.

Fig.\ \ref{fig:7}b shows the normal stress response for a fibrin gel polymerized this time at 22\textdegree C. Although the modulus $G$ of the gel does not change significantly, the mesh size of the gel is larger when the gel is polymerized at a lower temperature and hence the characteristic time scale $\tau$ is expected to be shorter (and the characteristic frequency higher). This means that if both gels are sheared at the same frequency, they are expected to show different Lissajous shapes. Indeed, the Lissajous curve of the 27\textdegree C gel sheared at $\nu=0.01$~Hz is similar to the Lissajous curve of the 22\textdegree C gel sheared at $\nu=0.1$~Hz. We observe excellent agreement between the data and the model over the entire range of applied oscillation frequencies both for fibrin gels polymerized at 22\textdegree C and 27\textdegree C.

\begin{figure}[htb]
	\includegraphics[width=8.5cm]{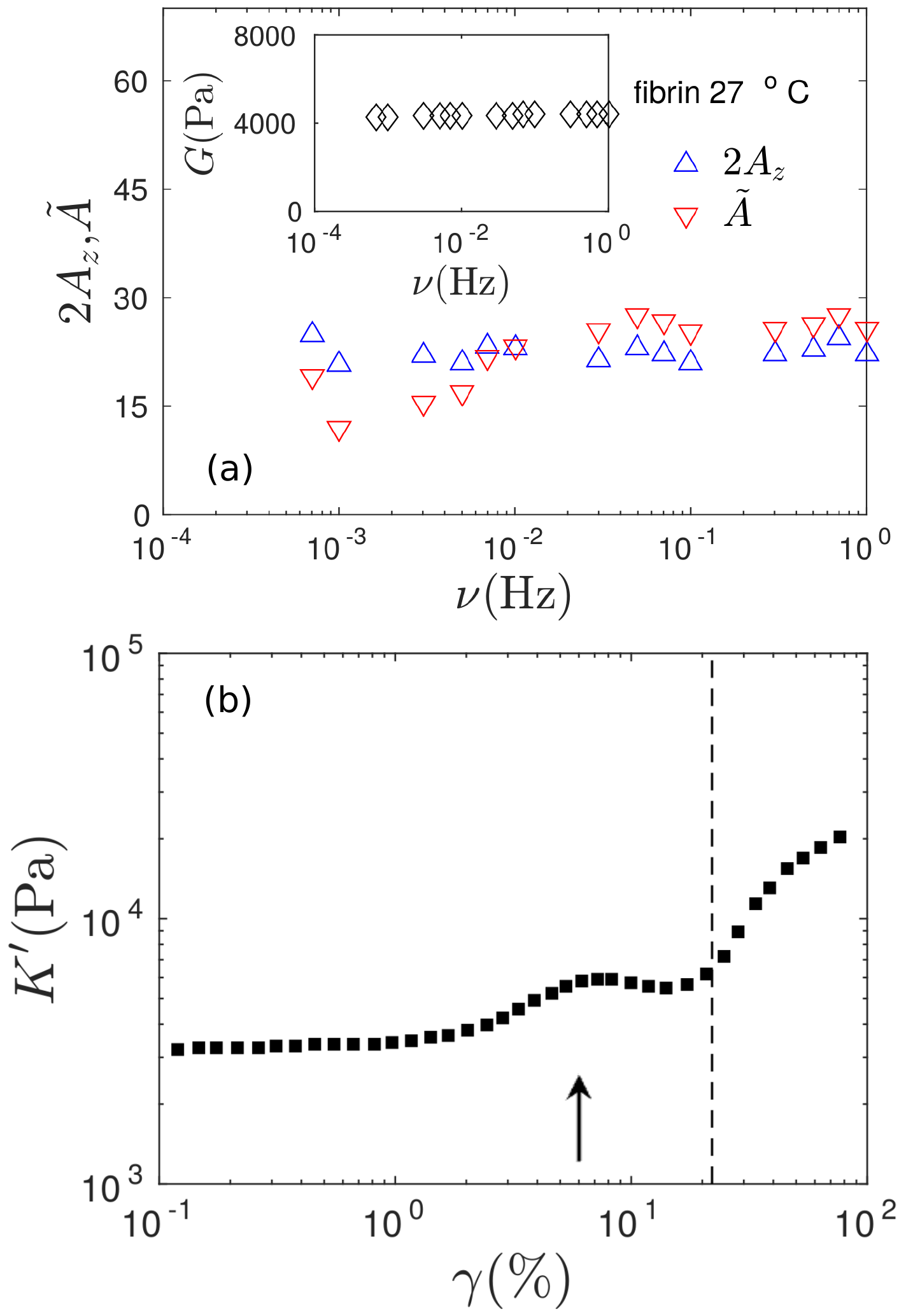}
	\caption{
		(color online) (a) Fitting parameters $2A_z$ and $\tilde{A}$ versus frequency for a fibrin gels polymerized at 27\textdegree C. The inset shows the shear modulus versus frequency, measured independently by rheology.
		(b) Differential storage modulus $K^{\prime}$ as a function of applied shear strain for a fibrin gel polymerized at 27\textdegree C. The arrow indicates the onset of nonlinearity at $\gamma_0 = 6.0\%$, calculated using Eq.\ \eqref{Az_gamma0}. The dashed line corresponds to an applied shear stress of 800~Pa, which is the applied stress in Figs.\ \ref{fig:6} and \ref{fig:7}.
		% Data: 170707K
	} 
	\label{fig:8}
\end{figure}

In Fig.\ \ref{fig:8}a, the resulting fitting parameters $2A_z$, $\tilde{A}$ and shear modulus $G$ are plotted as a function of frequency for a 27\textdegree C gel. The first of these parameters $2A_z$ is insensitive to the frequency of the oscillation, as expected from the model. The parameter $\tilde{A}$ is also insensitive to frequencies $\nu\gtrsim10^{-2}$. The observed deviation in $\tilde{A}$ for lower frequencies is to be expected, since the normal stress here is expected to be dominated by the axial stress $\sigma_{zz}$, which corresponds to the first term in Eq.\ \eqref{N1app}. Equivalently,  both $\mathcal{A}$ and $\mathcal{B}$ vanish in the low frequency limit, as can be seen in Fig.\ \ref{fig:NormStress1}. Thus, the fitting becomes increasingly independent of $\tilde{A}$ at low frequency, making the values of $\tilde{A}$ unreliable there. In practice, for $\nu\lesssim10^{-2}$, the fits in Fig.\ \ref{fig:7}a would be largely unchanged using the nearly constant values of $\tilde{A}$ obtained for $\nu\gtrsim10^{-2}$.
As a further test of our model, we note that the relative values of the parameters $2A_z$ and $\tilde{A}$ are roughly consistent with the prediction $\tilde A=2A_z+1$, particularly in the regime $\nu\gtrsim10^{-2}$, where both can be obtained reliably. This prediction is a consequence of the model in Sec.\ \ref{Model}, and does not depend on the specific stress calculations in Sec.\ \ref{C}.

In order to test the predictions of the semiflexible model in Sec.\ \ref{C}, we note that the values of $2A_z$ and $\tilde{A}$ are expected to vary with the onset strain $\gamma_0$, at which nonlinearity in the shear response appears. In Fig.\ \ref{fig:8}a, we see that $2A_z \simeq 22$, which implies a value of $\gamma_0\simeq0.06$ according to Eq.\ \eqref{Az_gamma0}. 
To verify if this is in agreement with experiments, we show in Fig.\ \ref{fig:8}b the differential or tangent storage modulus $K=d\sigma/d\gamma$ of a fibrin gel polymerised at 27\textdegree C. The arrow indicates 
$\gamma_0\simeq0.06$, which is in good agreement with the onset of nonlinearity in $K$. 

Thus, we find very good agreement overall with the main predictions of our model in Secs.\ \ref{Model} and \ref{C}. These predictions are, strictly speaking, derived in the low strain regime, corresponding to linear shear elasticity. We note that our experimental results in Figs.\ \ref{fig:6}-\ref{fig:7} were measured at a strain amplitude near 20\%, as indicated by the dashed line in Fig.\ \ref{fig:8}b. This is a level of strain below the point at which the response becomes strongly nonlinear: the differential modulus at $\gamma\simeq20\%$ is within approximately a factor of 2 of its linear value. In practice, it is difficult to measure normal stress accurately at lower strains over the full frequency range we study here, since such stress varies quadratically with strain: at a strain level indicated by the arrow in Fig.\ \ref{fig:8}b, the normal stresses would already be approximately a factor of 10 smaller, which would significantly reduce our ability to accurately probe the time dependence shown in Fig.\ \ref{fig:7}. Nevertheless, our results here provide significant support for both our general model of normal stresses in porous hydrogels in Sec.\ \ref{Model}, as well as the specific predictions in Sec.\ \ref{C} of the various stress components for semiflexible polymer gels. 

\section{Conclusions}
\sn{

Here, we have presented an extended derivation of the phenomenological model for normal stress relaxation that was introduced in Ref.\ \cite{DeCagny2016}. 
In the present work, we have also given a derivation of the key phenomenological parameters in the earlier model. 
We have done this for the specific case of semiflexible polymer networks \cite{MacKintosh1995,Gardel2004,Storm2005}.
While the phenomenological model presented in Sec.\ \ref{Model} should be more generally applicable to two-component gels, the derivation in Sec.\ \ref{C} is limited to networks of semiflexible chains with persistence length $\ell_p$ of order or larger than the mesh size $\xi$ of the network. 
Moreover, our derivation in Sec.\ \ref{C} assumes that the network deforms affinely. 

Our experimental results for fibrin gels are consistent with the phenomenological model in Sec.\ \ref{Model}, as well as the more specific predictions of the stress components in Sec.\ \ref{C}. Our calculation of stresses in Sec.\ \ref{C} shows that the phenomenological parameters $\tilde A$ and $A_z$ should depend on the network, but not on the frequency of oscillation. Importantly, our experiments demonstrate a nearly frequency-independent value of $A_z$, even though this was allowed to be a free parameter in the fits for various frequencies $\nu$, ranging from $\simeq10^{-3}-1$ Hz. While some variation in fit values for $\tilde A$ is observed at low frequencies $\lesssim10^{-2}$, this is a range where the normal stress becomes insensitive to $\tilde A$, as can be seen from Eq.\ \eqref{N1app} and the fact that both $\mathcal{A}$ and $\mathcal{B}$ decrease at low frequency $\omega\tau\lesssim1$ (Fig.\ \ref{fig:NormStress1}). 
In this low frequency regime, the Lissajous curves of normal vs shear stress in Fig.\ \ref{fig:7} reverse from concave up to concave down. 
Thus, the relaxation time $\tau$ in Eq.\ \eqref{tau} is expected to govern the sign of the apparent normal stress, independent of the microscopic details, including polymer flexibility. 

The basic mechanism determining the sign of normal stress is the porosity and relative motion of network and solvent, as sketched in Fig.\ \ref{fig:1}. 
Qualitatively, the torsion of the rheometer tends to squeeze out solvent and drive the network radially inward, as the sketch suggests. 
This is similar to other processes of syneresis, in which solvent can be expelled from a gel. 
But, it is important to note that our model does not suggest or require the macroscopic separation of solvent and network. 
In fact, with fixed boundary conditions of the network to the rheometer surfaces, only a small displacement of order the gap size $d$ or smaller is expected. 
This is consistent with the experimental observation that no solvent is irreversibly expelled in the course of an oscillatory shear. 
In principle, this mechanism of local relative motion of solvent and network can apply to any two-component gel with network and solvent. In practice, however, the relaxation time $\tau$ can become very long for small pores, making the gel effectively behave as incompressible single-component systems. 

Finally, it is worth noting the distinction between the phenomenology of a negative or inverse Poynting effect and the sign of $N_1$.
The Poynting effect refers to the elongation and corresponding axial compressive stress in a system subject to torsion. 
In the rheology of single-component polymer systems, for instance, this is regularly observed in the form of rod-climbing or the Weissenberg effect \cite{Larson1998} and can be directly attributed to a positive (first) normal stress difference $N_1=\sigma_{xx}-\sigma_{zz}$. 
But, this interpretation assumes an incompressible material. 
As shown in Ref.\ \cite{Janmey2007}, a wide range of biopolymer gels exhibit an inverse Poynting effect with tensile axial force. 
It was argued that this was due, in part, to the two-component nature of such hydrogels. 
This was confirmed in Ref.\ \cite{DeCagny2016} for fibrin gels, where it was also shown that even PAAm gels can exhibit an inverse Poynting effect on long enough time scales. 
However, both of these systems were shown to exhibit a positive or conventional Poynting effect on short time scales, where the systems effectively become incompressible, due to the viscous coupling of solvent and gel.
This behavior is consistent with a strictly positive normal stress difference $N_1=\sigma_{xx}-\sigma_{zz}$. 
This begs the question as to whether $N_1=\sigma_{xx}-\sigma_{zz}<0$ is possible. 
}

\section*{ACKNOWLEDGEMENTS}  M.V., B.E.V., and H.C.G.d.C. were supported by Stichting voor Fundamenteel Onderzoek der Materie (FOM), which is part of the Nederlandse Organisatie voor Wetenschappelijk Onderzoek (NWO). 
The authors thank A. J. Licup for useful comments. G.H.K. and F.C.M. acknowledge support from the FOM (Program Grant No. 143). F.C.M. was supported in part by the National Science Foundation (Grant No. PHY- 1427654).

%%%REFERENCES%%%
%\bibliographystyle{PRE} %the RSC's .bst file
\frenchspacing

\end{document}